\documentclass[hyper]{JHEP3} 

\JHEPspecialurl{http://jhep.sissa.it/JOURNAL/JHEP3.tar.gz}

\usepackage{epsfig,multicol,bbm}

\newcommand\fverb{\setbox\fverbbox=\hbox\bgroup\verb}
\newcommand\fverbdo{\egroup\medskip\noindent%
            \fbox{\unhbox\fverbbox}\ }
\newcommand\fverbit{\egroup\item[\fbox{\unhbox\fverbbox}]}
\newbox\fverbbox

\title{A new approach to calculate the gluon polarization}

\author{F.~Taghavi-Shahri$^{a}$, A.~Mirjalili$^{a,b}$ and M.~M.~Yazdanpanah$^{a,c}$\\
$^{a}$ School of Particles and Accelerators, Institute for Research
in Fundamental Sciences
(IPM) P.O. Box 19395-5531, Tehran, Iran\\
$^{b}$ Physics Department, Yazd University, Yazd, Iran\\
$^{c}$ Physics Department, Kerman Shahid Bahonar University, Kerman, Iran\\
    E-mail: ~\email {f\_taghavi@ipm.ir}, \email{Mirjalili@Mail.ipm.ir}, \email{myazdan@mail.ipm.ir}}

\received{\today}       
\accepted{\today}       

\abstract{We derive the Leading-Order master equation to extract
the polarized gluon distribution $ G(x,Q^2)=x \delta g(x,Q^2)$
from polarized proton structure function, $g^{p}_{1}(x,Q^{2})$. By
using a Laplace-transform technique, we solve the master equation
and derive the polarized gluon distribution inside the proton. The
test of accuracy which are based on our calculations with two
different method confirms that we achieve to the correct solution
for the polarized gluon distribution. We show that accurate
experimental knowledge of $g_1^{p}(x,Q^2)$ in a region of Bjorken
x and
 $Q^2$,  is all that is needed to determine the polarized
gluon distribution in that region. Therefore,  to determine the
gluon polarization $\frac{\delta g}{g}$  ,we only need to have
accurate experimental data on un-polarized and polarized structure
functions ($F_2^{p}(x,Q^2)$ and $g_1^{p}(x,Q^2)$).}

\keywords{Parton Model, Polarized gluon distribution, Laplace
transform}

\begin{document}


\section{Introduction}

The spin structure of the proton is one of the most challenging
open puzzles in Quantum Chromodynamics. One of the main questions
in particle and nuclear physics is: How is the proton spin built
up from its quark and gluon constituents? Now, we know that the
quark contribution to the spin of the proton is about 0.30. It is
determined precisely in a QCD fit to the polarized proton
structure function data,  $g_1^p(x,Q^2)$. Therefore at the present time, the role played by the gluons in the nucleon spin is the most challenging task. \\
The main goal of the experiments at HERMES, SMC, COMPASS and RHIC
spin physics programs is the measurement of the helicity
contribution of the gluons to the nucleon spin, $\Delta g$.
Experimentally this value is mainly accessible via two processes
in polarized DIS experiments. The first one is the production of
high $p_t$ hadron pairs with large transverse momentum, but these
processes have large background contributions from QCD Compton
processes and fragmentation that  one has to control. The second
one is open charmed meson production in photon-gluon fusion
process. The cross section of these two processes  is directly
related to the ratio of the polarized gluon density to the
un-polarized gluon density, $\frac{\delta g (x,Q^2)}{g(x,Q^2)}$
\cite{1,2}. Then, to measure the gluon contribution to the nucleon
spin with this approach, it is necessary to know the un-polarized
gluon distribution well. Polarized gluon measurements from deep
inelastic experiments
are summarized in Table 1.\\
In line with experiments, many NLO QCD global fits to the
inclusive $g_1^p(x,Q^2)$ data were used to extract the magnitude
of $\Delta g$ and the shape of $\delta g(x,Q^2)$ \cite{8,9,10}.
Unfortunately, these global fits are sensitive to the initial
assumption  for the polarized gluon distribution  and also to the
initial scale of $Q_0^{2}$. The shape of the polarized gluon
distribution extracted from different groups are not  identical
because of their different assumptions for the initial polarized
gluon distribution, initial $Q_0^{2}$ and also different
approaches for global fits. We would like to know how big the
gluon spin contribution is to the total spin of the proton and
look for a simple way to measure this value.
 This work can be considered as a proposal for
direct measurement of the
polarized gluon distribution inside the proton with more accuracy.\\
Recently an explicit expression for the un-polarized gluon
distribution function $G(x,Q^2)=x g(x,Q^2)$ in the proton in terms
of the proton structure function $F_2^p(x,Q^2)$ was derived by
using a Laplace-transform technique \cite{11,12,13}. Here, the
same procedure is used to derive the polarized gluon distribution
function, $G(x,Q^2)=x \delta g(x,Q^2)$, inside the proton. We
obtain an analytic solution for the polarized gluon distribution
in terms of the polarized proton structure function,
$g_1^p(x,Q^2)$. Thus we can calculate the polarized gluon
distribution inside the proton directly only by finding
$g_1^p(x,Q^2)$ over large kinematic range of $x$ and $Q^2$.  This
is the principal theoretical result in this paper.
\\
This paper is organized as follows. In section 2, we derive LO
master equation for extracting the polarized gluon distribution
 $G(x,Q^2)= x \delta g(x,Q^2)$ from
 polarized proton structure function, $g^{p}_{1}(x,Q^{2})$. Section 3 is
devoted to the solution of the master equation. In the last
section,  we  first perform  a global parametrization of
$g_1^{p}(x,Q^2)$ using all available experimental data and then we
calculate numerically the polarized gluon distribution. Our
conclusion is given in section 5.

\begin{table}
\begin{tabular}{|c|c|c|c|c|c|}
  \hline
  Experiment & process  & $<x_g>$ & $<\mu^2>$ &$\frac{\delta g(x_g, \mu^2)}{g(x_g, \mu^2)}$ & Reference\\
  \hline
   \hline
  HERMES & hadron pairs  & 0.17 &$\sim2 $& $0.41\pm0.18\pm0.03$& \cite{3} \\
  \hline
  HERMES & inclusive hadrons  & 0.22 & 1.35 & $0.071\pm0.034\pm0.105$ &\cite{2}\\
  \hline
  SMC & hadron pairs  & 0.07 &---  & $-0.20\pm0.28\pm0.10$ &\cite{4}\\
  \hline
  COMPASS & hadron pairs, $Q^2 < 1$ &0.085 &$ \sim 3 $& $0.016\pm0.058\pm0.055$ &\cite{5,6} \\
   \hline
  COMPASS & hadron pairs, $Q^2 > 1$  & 0.13 & ---& $0.06\pm0.31\pm0.06$ &\cite{5,6}\\
   \hline
  COMPASS & open charm & 0.15 & 13 & $-0.57\pm0.41\pm0.17$ &\cite{7}\\

  \hline
\end{tabular}
\caption{\label{label}Polarized gluon measurements from deep
inelastic experiments.}
\end{table}

\section{LO master equation to extract the polarized gluon distribution $ G(x,Q^2)=x \delta g(x,Q^2)$, using the
 polarized proton structure function, $g^{p}_{1}(x,Q^{2})$ }

The LO DGLAP equation, \cite{14,15,16} for the evolution of the
polarized proton structure function, $g^{p}_{1}(x,Q^{2})$   can be
written as (See Appendix A)

\begin{equation}
x \frac{\partial g_1^{  p}(x,Q^2)}{\partial\ln Q^2}  -
\frac{\alpha_s}{2\pi} x \int_{x}^1\frac{dz}{z}  g_1^{
p}(z,Q^2)\delta K_{qq}\left(\frac{x}{z}\right)
=x\frac{\alpha_s}{2\pi} \sum e_{i}^{2}
\int_{x}^1\frac{dz}{z^2}G(z,Q^2)\delta
K_{qg}\left(\frac{x}{z}\right),
\end{equation}
where $\delta K_{qq}(x)$ and $\delta K_{qg}(x)$ are the LO
polarized splitting functions and $\alpha_s$ is the renormalized
 running coupling constant. We introduce ${\cal G}_1^{
p}(x,Q^2)$ as

\begin{equation}
{\cal G}_1^{  p}(x,Q^2)=x \frac{\partial g_1^{
p}(x,Q^2)}{\partial\ln Q^2}  - \frac{\alpha_s}{2\pi} x
\int_{x}^1\frac{dz}{z}  g_1^{  p}(z,Q^2)\delta
K_{qq}\left(\frac{x}{z}\right).
\end{equation}
 Now we can write the master equation,  the DGLAP equation for the
$g^{p}_{1}(x,Q^{2})$,  as follows

\begin{equation}
gg(x,Q^2)=x \int_{x}^1\frac{dz}{z^2}G(z,Q^2)\delta
K_{qg}(\frac{x}{z}),
\end{equation}
where $gg(x,Q^2)=(\frac{\alpha_s}{2\pi}\sum e_{i}^{2} )^{-1}{\cal
G}_1^{  p}(x,Q^2)$. In the Eq. (2.2) and Eq. (2.3),  the LO
 $q\rightarrow q$ and $g\rightarrow q$ polarized splitting function
are given by \cite{17}

\begin{eqnarray}
\delta K_{qq}(x)=\frac{4}{3}((\frac{1+x^2}{1-x})_+ +\frac{3}{2}\delta (x-1)),\\
 \delta K_{qg}(x)=\frac{1}{2}(2x-1)=x-\frac{1}{2}.
\end{eqnarray}
To extract the polarized gluon distribution, $G(x,Q^2)= x \delta
g(x,Q^2)$ from the DGLAP equation, we should solve the master
equation (2.3) and find the polarized gluon distribution inside
the proton. This issue is the subject of the next section.

\section{Solution of the master equation  }
To solve the master equation (2.3), we follow the procedure that
was used by BDM \cite{11} and used the Laplace transformation to
solve this equation. Now we use the coordinate transformation as

\begin{equation}
v\equiv \ln(1/x),
\end{equation}
Then the functions $\hat{G}$, $\hat{K}_{qg}$, and $\hat{\cal G}$
in $v$-space are given by

\begin{eqnarray}
\hat{G}(v,Q^2) \equiv G(e^{-v},Q^2),\\
 \delta \hat K_{qg}(v,Q^2) \equiv \delta K_{qq}(e^{-v},Q^2),\\
\hat{\cal G} (v,Q^2)\equiv gg(e^{-v},Q^2).
\end{eqnarray}
Explicitly from Eq. (2.5), we have

\begin{equation}
\delta \hat{K}_{qg}(v,Q^2) = e^{-v}-\frac{1}{2}.
\end{equation}
Therefore

\begin{equation}
\hat{\cal G}(v,Q^2)=\int^v_0\hat G(w,Q^2)e^{-(v-w)}\delta
\hat  K_{qg}(v-w)\,dw \\
=\int^v_0\hat G(w,Q^2)\hat H(v-w)\, dw,
 \end{equation}

 where $w\equiv \ln(1/z)$ and $\hat H(v)$ is defined as

\begin{equation}
 \hat H(v)\equiv e^{-v} \delta \hat K_{qg}(v)\\
= e^{-2v} -\frac{1}{2} e^{-v}.
\end{equation}

If we take the laplace transform of Eq. (3.6), then we have

\begin{eqnarray}
{\cal L}[\hat{\cal G}(v,Q^2);s]={\cal L}[\int^v_0\hat G(w,Q^2)\hat
H(v-w)\, dw;s]\\
\Longrightarrow \hat{\cal G}(s,Q^2)=\hat G(s,Q^2)\times h(s).
\end{eqnarray}

In the above equation we used the following property for Laplace
transformation
\begin{equation}
{\cal L}^{-1}[F(s)G(s)]=\int^t_0 f(t-\tau)g(\tau)d\tau=\int^t_0
f(\tau)g(t-\tau)d\tau.
\end{equation}

So we have the polarized gluon distribution in s-space as

\begin{equation}
\hat G(s,Q^2)=\frac {\hat {\cal G}(s,Q^2)}{h(s)}.
\end{equation}

Now the polarized gluon distribution in v-space is given by

\begin{equation}
\hat G(v,Q^2)={\cal L}^{-1}[\hat {\cal
G}(s,Q^2)h(s)^{-1}]=\int^v_0\hat {\cal G}(w,Q^2)\hat J(v-w) dw,
\end{equation}

Where $\hat J(v) \equiv {\cal L}^{-1}[h(s)^{-1},v] $. The
calculation of $\hat J(v)$ by using the Eq. (3.7) and inverse
Laplace transform of $h(s)^{-1}$ for LO is straightforward and
given  in term of the Dirac delta function

\begin{equation}
\hat J(v)=4+6 \delta (v)+2 \delta^{'} (v).
\end{equation}

Using the Eq. (3.12) and Eq. (3.13) the polarized gluon
distribution in v-space is given by

\begin{equation}
\hat G(v,Q^2)=4 \int^v_0  \hat{\cal G}(w,Q^2)dw +6  \int^v_0
\delta (v-w) \hat{\cal G}(w,Q^2)dw+ 2 \int^v_0 \delta^{'}(v-w)
\hat{\cal G}(w,Q^2) dw.
\end{equation}

Using the following relation for Dirac delta function

\begin{eqnarray}
 \int^v_0
\delta (v-w) f(w)dw= f(v),\\
\int^v_0 \delta^{'}(v-w)f(w) dw=f^{'}(v).
\end{eqnarray}
we will have

\begin{equation}
\hat G(v,Q^2)= 4 \int^v_0  \hat{\cal G}(w,Q^2)dw +6 \hat{\cal
G}(v,Q^2)+ 2\frac {\partial \hat{\cal G}(v,Q^2)}{ \partial v}.
\end{equation}
Finally we have the polarized gluon distribution as

\begin{eqnarray}
G(x,Q^2)= 4 \int^1_x  gg(z,Q^2)\frac {dz}{z} +6gg(x,Q^2)-2( x\frac
{\partial gg(x,Q^2)}{ \partial
 x}),
\end{eqnarray}
where $gg(x,Q^2)=(\frac{\alpha_s}{2\pi}\sum e_{i}^{2} )^{-1}{\cal
G}_1^{  p}(x,Q^2)$ and ${\cal G}_1^{  p}(x,Q^2)$ is
given by Eq. (2.2). We use 4 massless quarks (u,d,s,c) in our calculation and then $\sum e_{i}^{2}=\frac {10}{9}$. \\
To calculate the right hand of the Eq. (3.18), we have to have an
analytic function for $g_{1}^p(x,Q^2)$. So in the next section we
try to do the global parametrization of $g_1^{p}(x,Q^2)$ using all
available experimental data and find this analytic function and
then we calculate the polarized gluon distribution function
numerically.


\section{Numerical results }
In this section we intend to  use Eq. (3.18) to extract the
polarized gluon distribution inside the proton.

\subsection{Global parametrization of $g_1^{p}(x,Q^2)$ using all available experimental data }

We have parameterized the polarized proton structure function,
$g_1^{  p}(x,Q^2)$ in $0<x<1$ as

\begin{equation}
xg_1^{  p}(x,Q^2)= x^a (1 - x)^b \frac {x_p  g_p}{(1-x_p)^b
x_p^a}(1+A(Q^2)Ln[\frac{x_p(1-x)}{x(1-x_p)}]).
\end{equation}
Here $x_P=0.234$ is  an approximate fixed point observed in the
data where the curves for different $Q^2$ cross. At that point,
$\partial g_1^{ p}(x_P,Q^2)/\partial \ln Q^2\approx 0$ for all
$Q^2$;  $g_P=g_1^{ p}(x_P,Q^2)=0.241$ is the common value of
$g_1^{  p}$. We used all available experimental data for polarized
proton structure function from E143, SMC, HERMES 2006 and COMPASS
2009   \cite{18,19,20,21,22}. More accurate global fit with more
data,  over large kinematic ranges of $x$ and $Q^2$ will lead to
the precisely determination of the polarized gluon distribution
function. The $Q^2$ dependence of $g_1^{ p}(x,Q^2)$ in our global
fit is given by

\begin{equation}
    A(Q^2)=a_0+a_1\ln Q^2 +a_2\ln^2 Q^2. \\
\end{equation}
The fitted quantities  are tabulated in Table 2.

\begin{table}
\begin{center}
\begin{tabular}{|c|c|}
  \hline
  $Parameters$ & $valuse$   \\
  \hline
   \hline
  $a_0$ & $-0.0287\pm0.1463$  \\
  \hline
  $a_1$ & $0.1253 \pm 0.0671$ \\
  \hline
  $a_2$ & $-0.0092 \pm 0.0145$ \\
  \hline
    a & $1.2624  \pm 0.0979$ \\
  \hline
  b & $2.227\pm 0.4745$  \\
   \hline
  \hline
$\chi^2$(Goodness of fit) & 0.984  \\
  \hline
\end{tabular}
\caption{\label{label} Global fit parameters obtained by fitting
the Eq. (4.1) over the experimental data .}
 \end{center}
\end{table}
We use the global fit in Eq. (4.1) and depict in Fig. 1 and Fig. 2
the polarized proton structure function for some values of $Q^2$
and compared them with experimental data. The comparison indicates
that the global fit works well.
\begin{figure}[htp]
\centerline{\begin{tabular}{cc}
\includegraphics[width=10 cm]{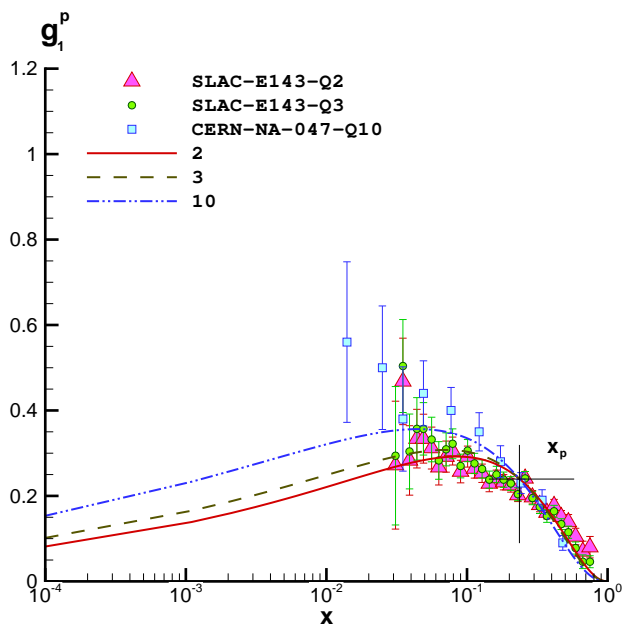}
\end{tabular}}
 \caption{\footnotesize  Polarized proton
structure function,  $ g_{1}^{p}$  for some values of $Q^2$
extracted from global fit. Experimental data are from
\cite{18,22}. The data cover the kinematic
 regions $0.0041 < x < 0.9$ and
 $0.18 GeV^2 < Q^2 < 20 GeV^2$. }
\end{figure}

\begin{figure}[htp]
\centerline{\begin{tabular}{cc}
\includegraphics[width=8cm]{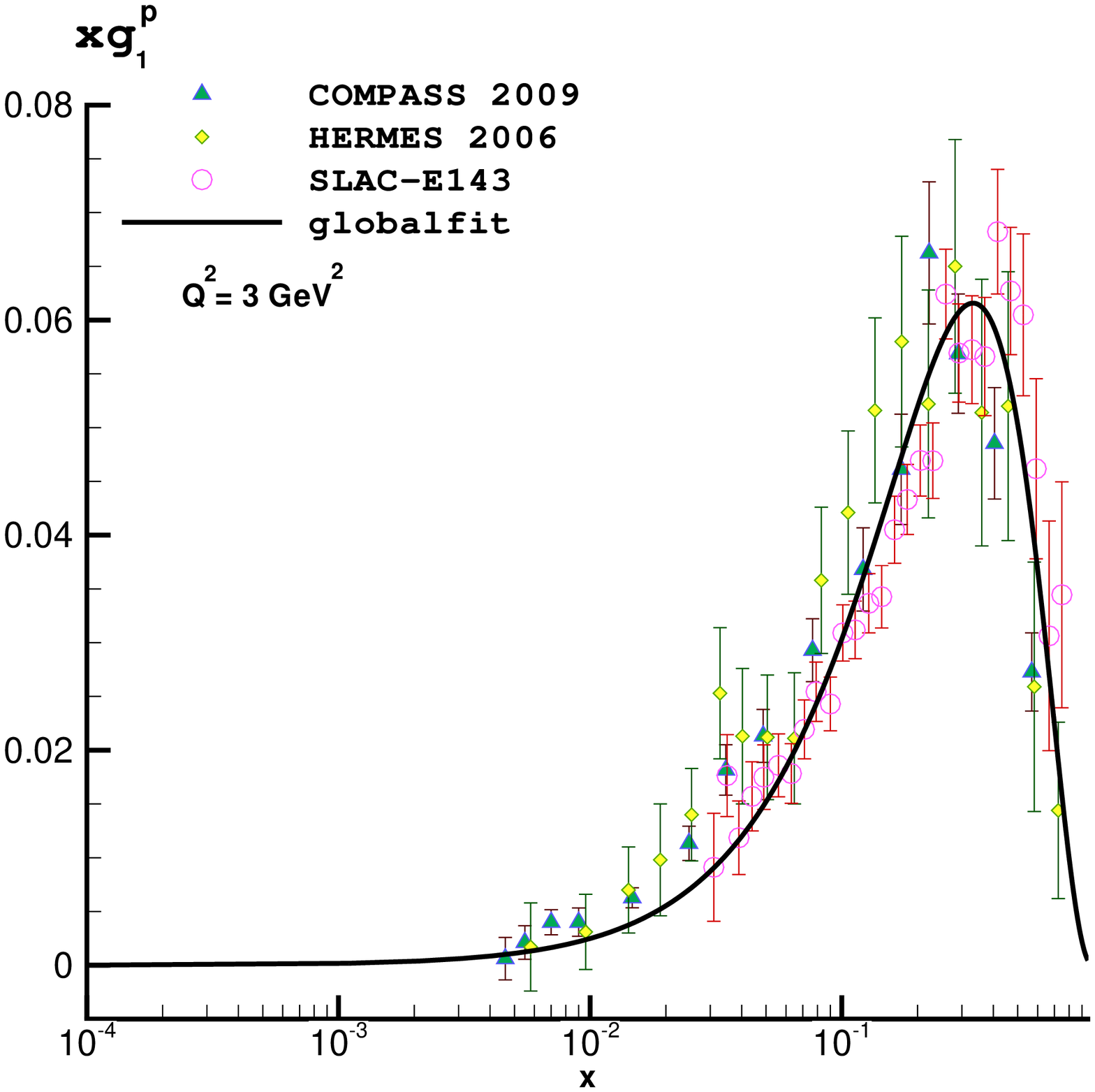}
\includegraphics[width=8 cm]{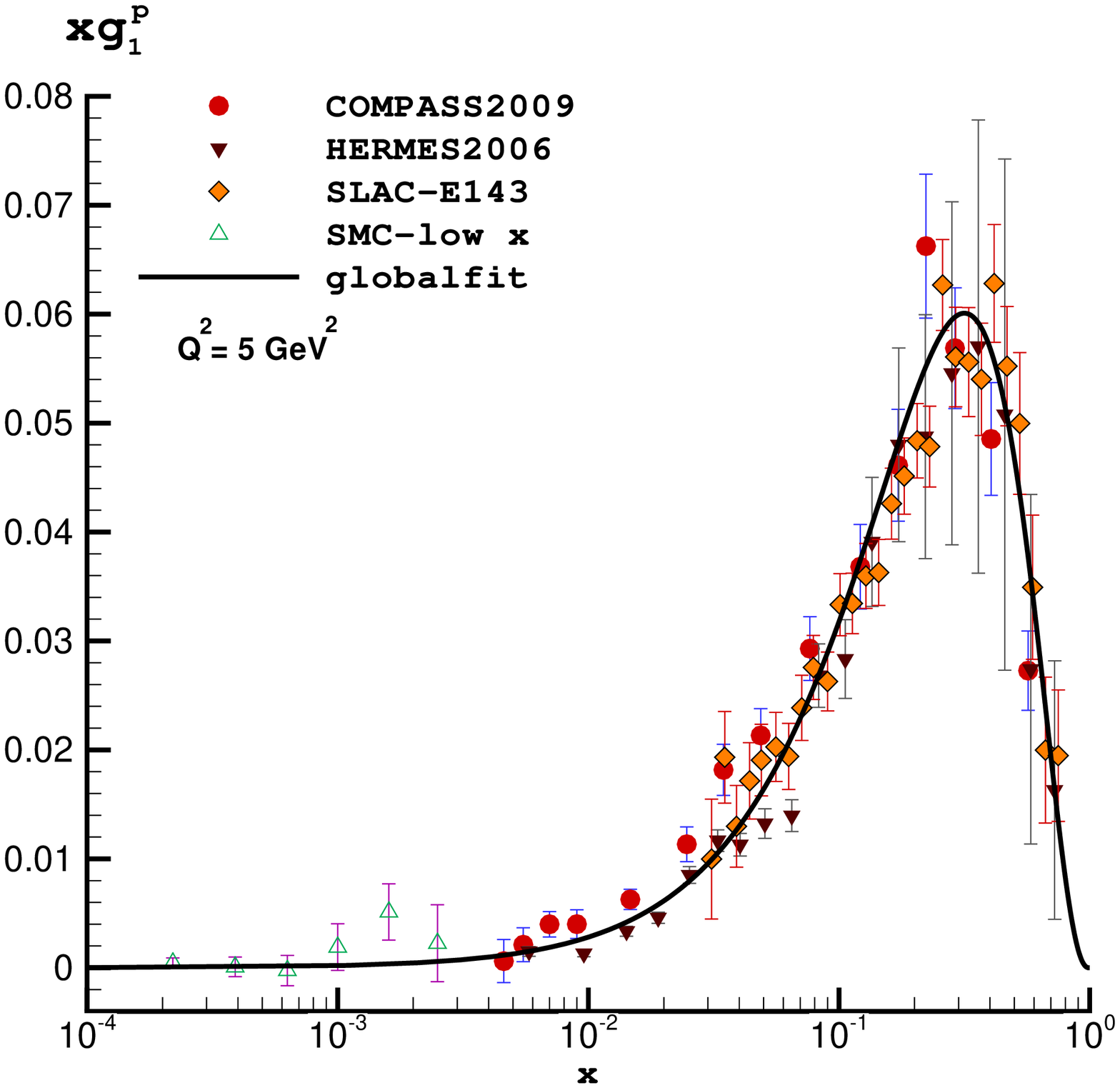}
\end{tabular}}
 \caption{\footnotesize  Polarized proton
structure function, $x g_{1}^{p}$ at $Q^2=3 GeV^2$ and
 $Q^2=5 GeV^2$ extracted from global fit.
Experimental data are from \cite{18,19,20,21,22,27}. }
\end{figure}

\newpage

\subsection{Numerical results for the polarized gluon distribution}

In this section we will briefly describe the numerical results for
polarized gluon distribution by using the analytic solution, Eq.
(3.18). Finally we will compare our result for $x \delta g(x,Q^2)$
with those from $AAC'08$, $DSSV'08$ and $LSS'06$ global fits
\cite{23,24,25}.\\
In this calculation, we use the LO approximation of
$\alpha_s(Q^2)$ which is defined in \cite{17}

\begin{eqnarray}
\alpha_s(Q^2) =\frac{4 \pi}{\beta_{0}\ln(Q^2/\Lambda^2)},
\\
\beta_{0}= 11-{2\over3}n_f,
\end{eqnarray}

with $n_f=5$ and $\Lambda_5=146$ MeV for $Q>4.5$ GeV, $n_f=4$ and
$\Lambda_4=192$ MeV for 1.3 GeV $<Q\leq 4.5$ GeV, and $n_f=3$ and
$\Lambda_3=221$ MeV for $Q<1.3$ GeV. These values have been used
in $CTEQ5L$ \cite{26} and also used to extract the analytic
unpolarized gluon distribution in \cite{11}.\\
Now by following these two steps we calculate the polarized gluon
distribution:

\begin{itemize}
\item Calculating $ gg(x,Q^2)$ by using the Eq. (2.2): $gg(x,Q^2)=(\frac{\alpha_s}{2\pi}\sum e_{i}^{2} )^{-1}{\cal
G}_1^{  p}(x,Q^2)$. The convolution integrate  in Eq. (2.2) with
plus prescription,$( )_+$,  can be easily calculated using
\cite{17}

\begin{equation}
\int^1_x \frac{dy}{y} f(\frac {x}{y})_+ g(y)=\int^1_x \frac{dy}{y}
f(\frac {x}{y})[g(y)-\frac {x}{y} g(x)]-g(x) \int^x_0 dy f(y)
\end{equation}

\item By using the Eq. (3.18), we can extract numerically the polarized gluon distribution inside the proton
.
\end{itemize}

Our result for polarized gluon distribution inside the proton is
shown in Fig.3 (left) and  compared with some  global fits.

\subsection{How to test the accuracy of the procedure ?}

An independent method of checking the numerical accuracy of the
entire procedure for extracting the polarized gluon distribution
is to go back to the original DGLAP equation from which we
started, Eq. (2.3), i.e.,

\begin{eqnarray}
gg(x,Q^2)=x \int_{x}^1\frac{dz}{z^2}G(z,Q^2)\delta
K_{qg}(\frac{x}{z}),\nonumber
 \end{eqnarray}
 and numerically integrate its right hand side, which
depends on our numerical solution . We then compare it with the
$gg(x,Q^2)$, which is independently known and arising from our
global fit for the polarized proton structure function,
$g_{1}^p(x,Q^2)$, based on its relation to ${\cal G}_1^{
p}(x,Q^2)$ in Eq. (2.2). In Fig.3 (right) we plot the results of
these different methods. Test of accuracy shows that our solution
for polarized gluon distribution is correct (Eq. (3.18)). Of
course, we should note that for using this solution for polarized
gluon distribution, we need to do a global fit with experimental
data for $g_1^p(x,Q^2)$. We do not have enough experimental data
for small x ($x<10^{-2}$) also for large x ($x>0.7$), therefore we
need to have $g_1^p(x,Q^2)$ at these regions to be sure about the
shape of polarized gluon distribution and future experiments
should be focused on measuring the polarized structure
function in these two regions.\\
The gluon contribution to the spin of the proton in our
calculation is
\begin{equation}
\Delta g(Q^2=5 GeV^2)=\int^{1}_{0.001} \delta g(x,Q^2=5
GeV^2)dx=0.43
\end{equation}
This value indicates that   the gluon contribution to the proton
spin  is considerable and compatible with other phenomenological
models \cite{23,24,25}.

\begin{figure}[htp]
\centerline{\begin{tabular}{cc}
\includegraphics[width=8cm]{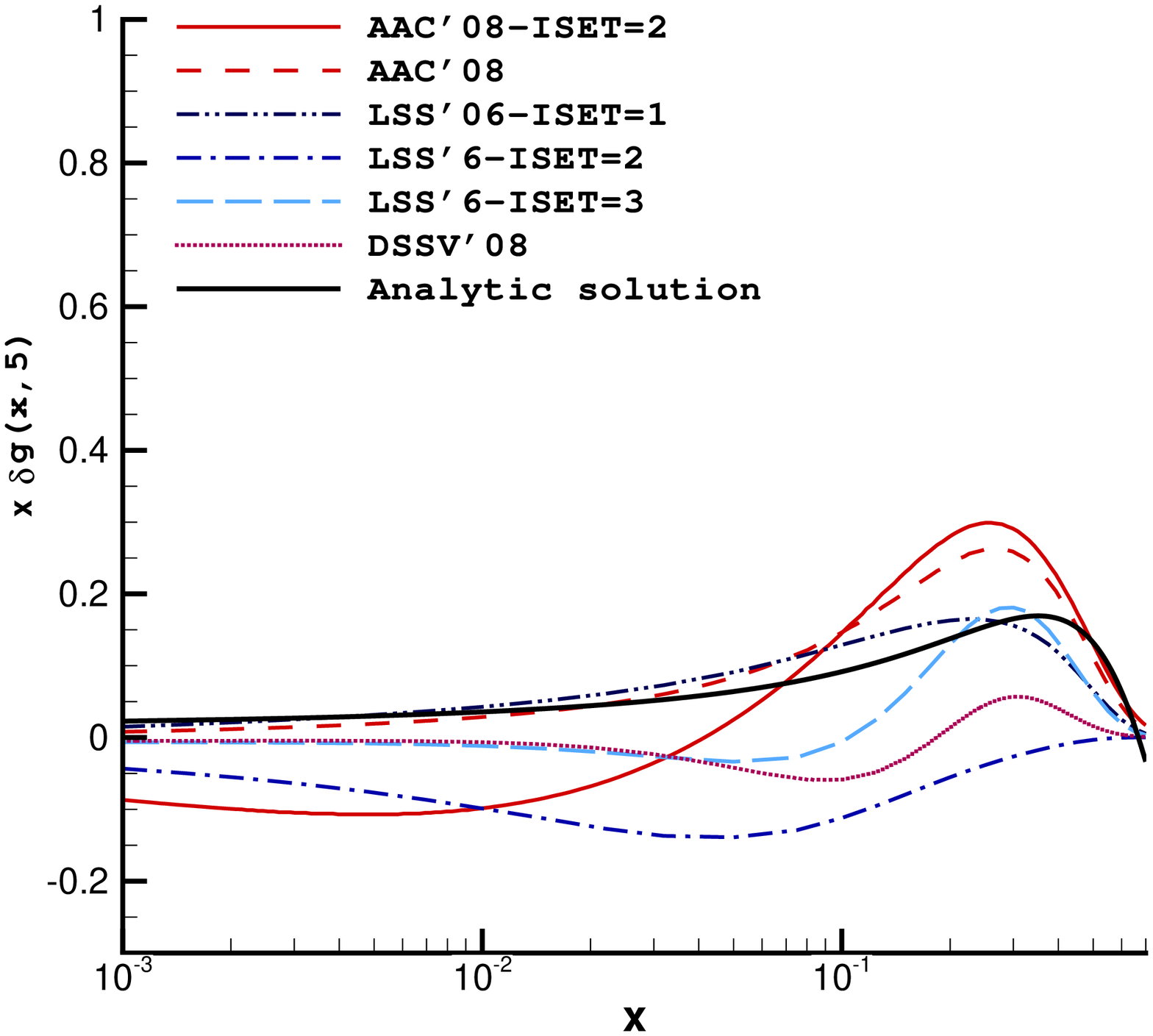}
\includegraphics[width=8cm]{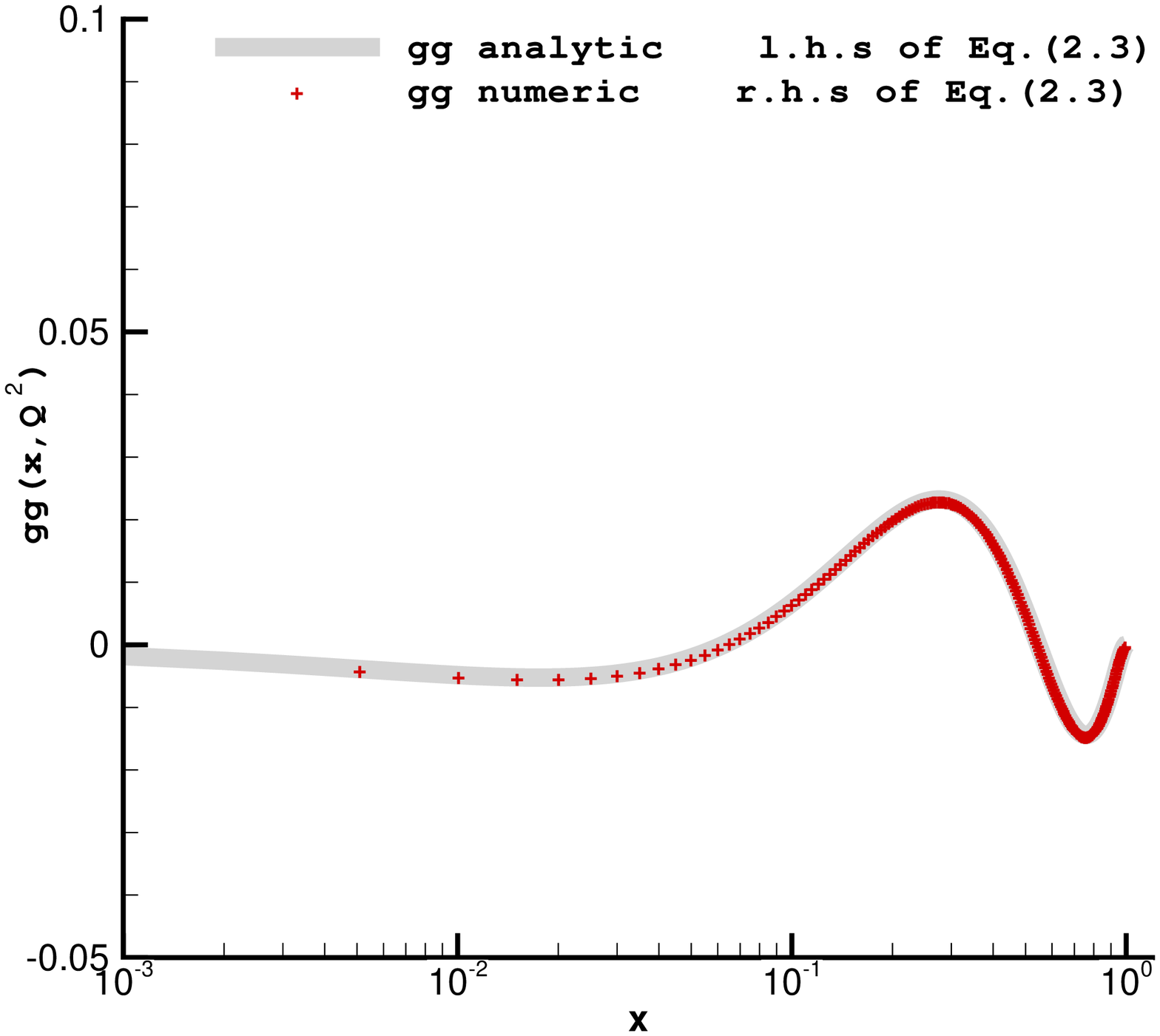}
\end{tabular}}
 \caption{\footnotesize {\it{Left:}} Polarized gluon
distribution function, $x \delta g(x,Q^2)$ at $Q^2=5 GeV^2$ and
comparison with other global fits.   {\it{Right:}} Test of
accuracy.}
\end{figure}

\newpage

\section{Conclusion}

In this paper we tried to find an analytic solution for the DGLAP
equation for polarized  structure function and find the polarized
gluon distribution inside the proton. Our solution is model
independent and it is free of any parameters, initial input
densities for solving the DGLAP equations and also free of the
initial scale of $Q_{0}^2$. It is only dependent on  finding an
analytic function for $g_{1}^p(x,Q^2)$    by using the
experimental data. Our results suggest that the precise
measurement of experimental values for  $g_{1}^p(x,Q^2)$ over
large kinematic ranges of $x$ and  $Q^2$  can predict directly the
polarized gluon distribution more accurately.

\section*{Acknowledgments}
Authors are indebted to the institute for research in fundamental
science (IPM) for their hospitality whilst  this research was
performed. We would like to thank  G.Altarelli for his careful
reading of the manuscript and for the productive discussions. We
are grateful to  M. Block for his useful suggestions, discussions
and critical remarks. The authors are indebted to R. Sassot  for
giving us his useful and constructive comments.

\newpage

\appendix

\section {The DGLAP equation for the polarized proton structure function}

In this part, we want to prove the Eq. (2.1).

\begin{equation}
x \frac{\partial g_1^{  p}(x,Q^2)}{\partial\ln Q^2}  -
\frac{\alpha_s}{2\pi} x \int_{x}^1\frac{dz}{z}  g_1^{
p}(z,Q^2)\delta K_{qq}\left(\frac{x}{z}\right)
=x\frac{\alpha_s}{2\pi} \sum e_{i}^{2}
\int_{x}^1\frac{dz}{z^2}G(z,Q^2)\delta
K_{qg}\left(\frac{x}{z}\right),
\end{equation}

We begin with the  DGLAP equations for the polarized parton
distribution functions for quark and ani-quark sectors

\begin{eqnarray}
\frac{\partial \delta q_i}{\partial t}(x,t)=\frac{\alpha_s}{2\pi}[
\int_{x}^1\frac{dz}{z}   \delta q_i(z,t) \delta
K_{qq}\left(\frac{x}{z}\right)+ \int_{x}^1\frac{dz}{z} \delta
g(z,Q^2)\delta K_{qg}\left(\frac{x}{z}\right)],\\
\frac{\partial \bar {\delta q_i}}{\partial
t}(x,t)=\frac{\alpha_s}{2\pi}[ \int_{x}^1\frac{dz}{z}  \bar
{\delta q_i}(z,t) \delta K_{\bar q \bar
q}\left(\frac{x}{z}\right)+ \int_{x}^1\frac{dz}{z} \delta
g(z,Q^2)\delta K_{\bar qg}\left(\frac{x}{z}\right)],
\end{eqnarray}

where  $t=Ln Q^2$ and we used the following properties  for the
polarized splitting functions \cite{17}:

\begin{eqnarray}
\delta K_{qq}=\delta K_{\bar q \bar q},\\
\delta K_{qg}=\delta K_{\bar q g},
\end{eqnarray}

After summation of the Eq. (A.2) and Eq. (A.3) and multiply both
side by $\frac {1}{2}\sum_{i=1}^{n_f}e_i^2 x$, we have

\begin{equation}
x \frac{\partial g_1^{  p}(x,Q^2)}{\partial\ln Q^2}  =
\frac{\alpha_s}{2\pi} x \int_{x}^1\frac{dz}{z}  g_1^{
p}(z,Q^2)\delta K_{qq}\left(\frac{x}{z}\right)
+x\frac{\alpha_s}{2\pi} \sum e_{i}^{2}
\int_{x}^1\frac{dz}{z^2}G(z,Q^2)\delta
K_{qg}\left(\frac{x}{z}\right),
\end{equation}

In the Eq. (A.6) we have $ G(z,Q^2)=z \delta g(z,Q^2)$ and $g_1^{
p}(x,Q^2)$ is the LO polarized spin structure function for proton

\begin{equation}
g_1^{  p}(x,Q^2)= \frac {1}{2 }\sum_{i=1}^{n_f}e_i^2 [\delta
q_i(x,Q^2)+ \delta\bar{q}_i(x,Q^2)].
\end{equation}

Thus we achieved to our desired result in Eq. (2.1):

\begin{equation}
x \frac{\partial g_1^{  p}(x,Q^2)}{\partial\ln Q^2} -
\frac{\alpha_s}{2\pi} x \int_{x}^1\frac{dz}{z}  g_1^{
p}(z,Q^2)\delta K_{qq}\left(\frac{x}{z}\right)
=x\frac{\alpha_s}{2\pi} \sum e_{i}^{2}
\int_{x}^1\frac{dz}{z^2}G(z,Q^2)\delta
K_{qg}\left(\frac{x}{z}\right),
\end{equation}

\end{document}